\documentclass[12pt,a4paper]{article}
\usepackage[T2A]{fontenc}
\usepackage[utf8]{inputenc}
\usepackage[english]{babel}
\usepackage{amssymb}
\usepackage{amsmath}
\usepackage{graphicx}
\usepackage[small,sc,center]{titlesec}
\usepackage{indentfirst}
\usepackage{siunitx}
\usepackage[labelsep=period]{caption}
\usepackage{caption}

\begin{document}

\begin{center}
	\textbf{The impact of metallicity on the period--luminosity relation of Mira variables}

	\vskip 3mm
	\textbf{Yu. A. Fadeyev\footnote{E--mail: fadeyev@inasan.ru}}

	\textit{Institute of Astronomy, Russian Academy of Sciences,
    		Pyatnitskaya ul. 48, Moscow, 119017 Russia} \\

	Received July 28, 2025; revised July 28, 2025; accepted August 1, 2025
\end{center}

\textbf{Abstract} ---
Evolution of stars with initial masses $M_\mathrm{ZAMS}=1.1M_\odot$, $1.3M_\odot$, $1.5M_\odot$
and relative mass abundances of metals $Z=0.006$ and 0.02 was computed from the main sequence
up to the final AGB stage.
Selected models of evolutionary sequences were used as initial conditions for solution of
the equations of hydrodynamics describing pulsations of red giants, whereas for each
evolutionary sequence of Mira variables pulsating in the fundamental mode we determined
the theoretical period--luminosity relation.
A change in the metal abundance is shown to substantially affect the period--luminosity relation
because of significant growth of the slope with decreasing $Z$.
In particular, Mira variables of the LMC ($Z=0.006$) are brighter by 0.2 -- 0.5 mag than
galactic Mira variables ($Z=0.02$) with same pulsation periods.
The low boundary of fundamental mode pulsations changes from $\Pi\approx 70$~dat for $Z=0.02$
to $\Pi\approx 120$~day for $Z=0.006$.


\section*{introduction}

The correlation between the period of light variations and bolometric luminosity of
long period Mira variables was found by Glass and Lloyd~Evans (1981) based on infrared
observations of red giants in the Large Magellanic Cloud (LMC).
This dependence was almost immediately recognized as a new independent anchor of the
interstellar distance scale no less reliable than the former period--luminosity relation
of classical Cepheids (Glass and Feast 1982; Feast 1984).
In the following years owing to infrared observations of red giants of the LMC
the period--luminosity relation underwent further corrections
(Feast et al. 1989; Hughes 1989; Hughes and Wood 1990; Reid et al. 1995;
Soszy\'nski et al.  2007; Whitelock et al. 2008; Yuan et al. 2017; Iwanek et al. 2021)
so that its potential was substantially enhanced.
At the same time the limits of applicability of this relation still remain unclear for
Mira variables with composition different from that of LMC stars (Feast 1996).
Significant difficulties appear due to insufficiently high precision of trigonometric
parallax measurements and thereby the less reliable period--luminosity relation determined
for galactic Mira variables (Andriantsaralaza et al. 2022; Sanders 2023).

Pulsating Mira--type variables belong to stars of the asymptotic giant branch (AGB)
so that the correlation between their period and magnitude is due to evolutionary
changes in the radius and luminosity (Vassiliadis and Wood 1993).
However, one should bear in mind that the evolutionary phase of Mira variables
is characterized by cyclic variations of radius and luminosity on the thermal time--scale
of the stellar envelope ($\sim 10^3$~yr) as well as on the more extended time intervals
$\sim 10^5$~yr due to thermal flashes in the helium--burning shell source
(Wood and Zarro 1981).
Therefore, the theoretical period--luminosity relation of Mira variables can be determined
only from the consistent stellar evolution and nonlinear stellar pulsation calculations.
No studies of this problem have been done till recent time and first results were presented
by the author where he considered the evolutionary and hydrodynamic models of Mira variables
with the solar metal abundance $Z=0.014$ (Fadeyev 2023) and with the metal abundance of LMC
stars $Z=0.006$ (Fadeyev 2024).
One of the preliminary conclusions of these works is that the slope of the
period--luminosity relation seems to increase with decreasing $Z$.

The abundance of metals (i.e. the elements heavier than helium) in the stellar gas plays
an important role in both the evolution and radial pulsation of the star because of
a direct impact on the gas density and opacity as well as on the sound speed.
The goal of the present work is to determine the period--luminosity relations of
Mira variables with the initial mass on the main sequence $M_\mathrm{ZAMS}=1.1M_\odot$,
$1.3M_\odot$ and $1.5M_\odot$ for two different metallicities $Z=0.006$ and $Z=0.02$.
To calculate stellar evolution we employed the MESA code version r15140
(Paxton et al. 2019), whereas the grid of thermonuclear reactions and the parameters
of the stellar evolution theory are the same as in our preceding works (Fadeyev 2023; 2024).
Solution of the equations of radiation hydrodynamics and time--dependent convection
describing radial stellar oscillations was obtained for the initial conditions
calculated for selected models of evolutionary sequences.
As a result, for all evolutionary sequences characterized by the metal abundance $Z$
and initial mass $M_\mathrm{ZAMS}$ we obtained  a number of period--luminosity relations.

\section*{theoretical period--luminosity relations}

For determination of the theoretical period--luminosity relations we used the
hydrodynamic models that satisfy the two following conditions.
First, we consider the models of Mira variables pulsating in the fundamental mode.
Second, we selected hydrodynamic models with fairly regular limiting amplitude oscillations
so that the error of the period estimate is less than one percent.
The results presented below were obtained from calculations of somewhat more than a hundred
hydrodynamic models with periods from 70 to 400 day.

\begin{figure}
 \centering
 \includegraphics[width=0.70\columnwidth,clip]{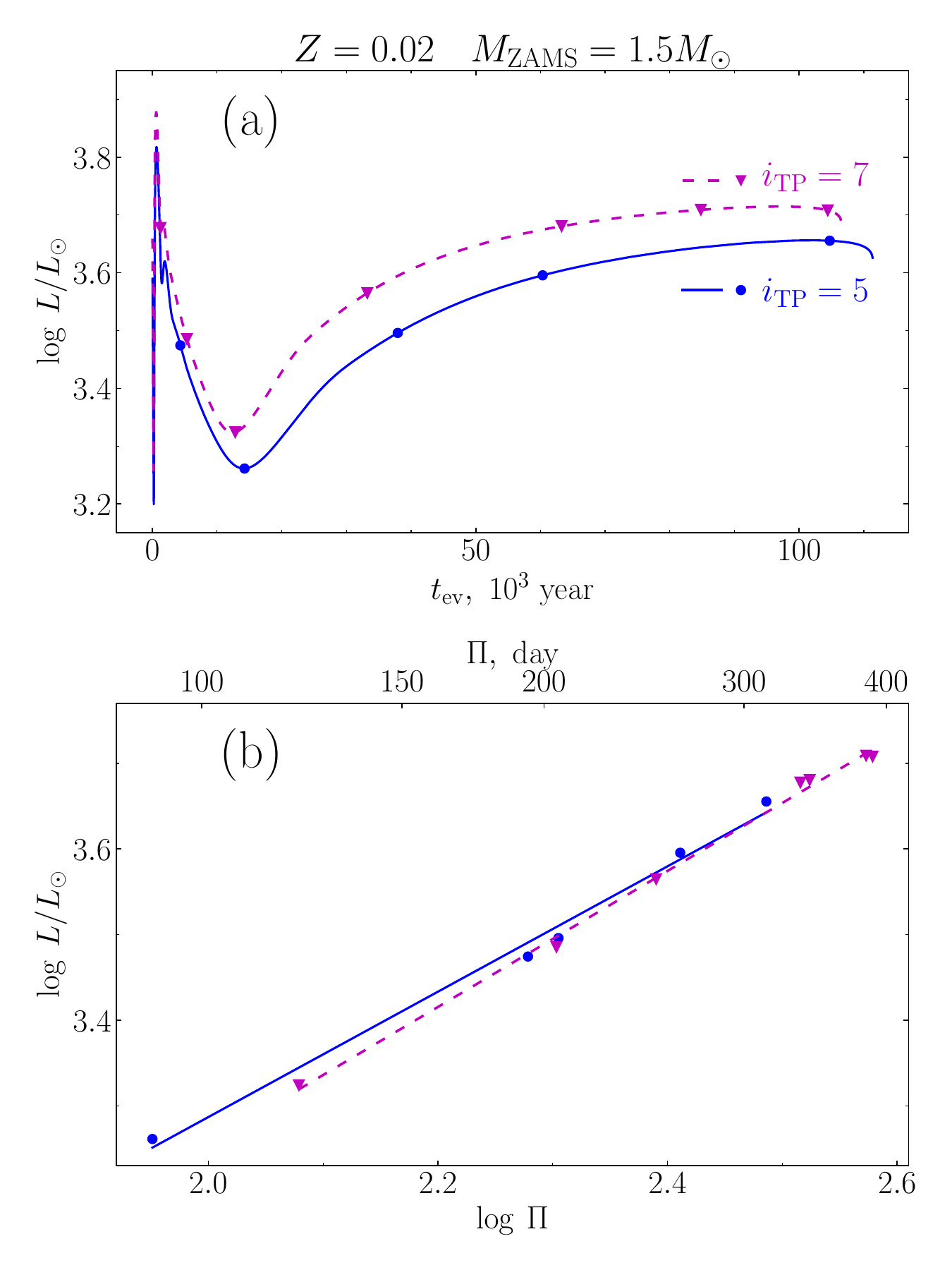}
 \caption{(a) -- Variations of stellar luminosity in the evolutionary sequence
 $Z=0.02$, $M_\mathrm{ZAMS}=1.5M_\odot$ after the 5--th (solid line) and 7--th
 (dashed line) thermal flashes. Circles and triangles indicate evolutionary models
 used as initial conditions in calculations of stellar pulsations.
 (b) -- Hydrodynamic models and regression lines after  the 5--th and 7--th
 thermal flashes on the period--luminosity diagram.}
\label{fig1}
\end{figure}

To better understand the origin of the correlation between the pulsation period $\Pi$ and
luminosity $L$ of Mira variables let us consider the evolution of the AGB star
between two consecutive thermal flashes.
Fig.~\ref{fig1}a shows the plots of the evolutionary variations of the stellar luminosity
after the fifth ($i_\mathrm{TP}=5$) and seventh ($i_\mathrm{TP}=7$) thermal flashes in
the evolutionary sequence $Z=0.02$, $M_\mathrm{ZAMS}=1.5M_\odot$.
For the sake of convenience the evolution time $t_\mathrm{ev}$ is set to zero at maximum
luminosity of the helium--burning shell source $L_\mathrm{He}$ and each plot extends to the
next maximum of $L_\mathrm{He}$.
The circles ($i_\mathrm{TP}=5$) and triangles ($i_\mathrm{TP}=7$) mark the evolutionary
models used as initial conditions for calculation of stellar pulsations.
In Fig.~\ref{fig1}b the same symbols show the mean stellar luminosity and pulsation period
of hydrodynamic models in the diagram period--luminosity.
As can be seen from the plots, during each cycle of thermal instability
the evolutionary changes of $\log L$ and $\log\Pi$
are fitted by the linear dependence with good accuracy.
At the same time the regression lines fitting evolutionary changes of $\log L$ and $\log\Pi$
shift from one to another with increasing $i_\mathrm{TP}$ due to evolutionary changes in
the carbon core mass and stellar luminosity.
In general for the evolutionary sequence
these differences lead to scatter of points on the period--luminosity diagram.

\begin{figure}
 \centering
 \includegraphics[width=0.70\columnwidth,clip]{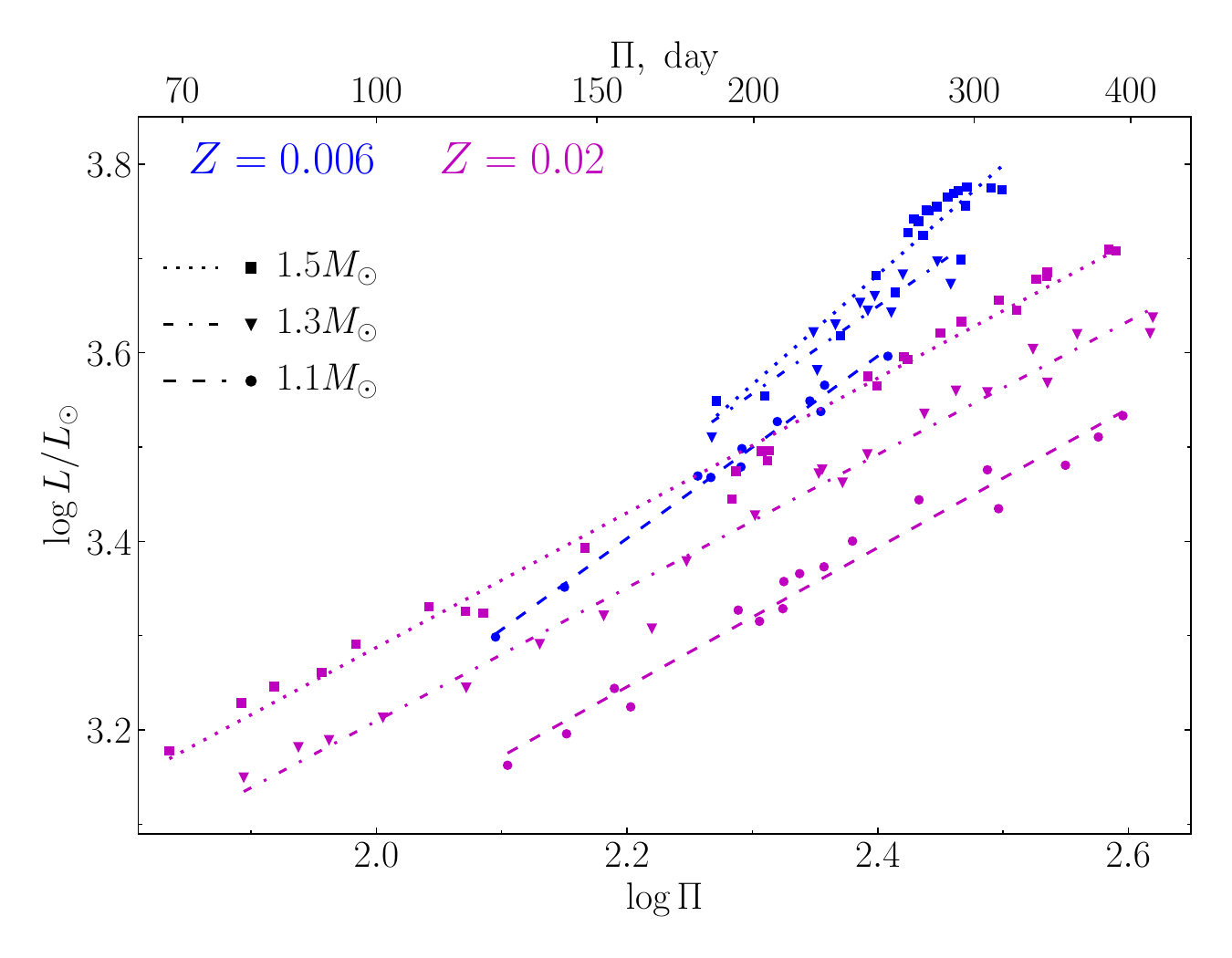}
 \caption{Hydrodynamic models of Mira variables of evolutionary sequences
 $M_\mathrm{ZAMS}=1.5M_\odot$ (squares), $1.3M_\odot$ (triangles) and  $1.1M_\odot$ (cicrcles)
 on the period--luminosity diagram for metal abundance $Z=0.006$ и $Z=0.02$.
 Linear fits of the period--luminosity relations are shown by dashed, dashed--dotted and
 dotted lines.}
\label{fig2}
\end{figure}

All results of evolutionary and hydrodynamic calculations carried out in the present study
are displayed in Fig~\ref{fig2}, where the circles, triangles and squares represent
hydrodynamic models of evolutionary sequences with initial masses $M_\mathrm{ZAMS}=1.1M_\odot$,
$1.3M_\odot$ and $1.5M_\odot$, respectively,
whereas the dashed, dashed--dotted and dotted lines are the corresponding regression lines.
These plots allow us to conclude the following.
First, the period--luminosity relations show the notable increase of the slope with decreasing
metal abundance.
Second, the decrease in metallicity is accompanied by increasing luminosity of pulsating
red giants.
For example, in the pulsation period range from 200 to 300 day the Mira variables of the
LMC ($Z=0.006$) are brighter than their galactic counterparts ($Z=0.02$) by 0.2 to 0.5 mag.
Third, the lower boundary of the fundamental mode oscillations moves to longer periods
with decreasing $Z$ from $\Pi\approx 70$~day for Galactic Mira variables to $\Pi\approx 120$~day
for Mira variables of the LMC.
Therefore, the most of LMC red giants with periods shorter
than 120 day are the first--overtone pulsators.

The main characteristics of period--luminosity relations are listed in Table~\ref{tabl1},
where $k_L$ and $a_0$ are the coefficients in the relation
\begin{equation*}
 \log L/L_\odot = k_L \log\Pi + a_0 ,
\end{equation*}
$s_k$ is the standard error of the estimated slope $k_L$,
$\Pi_a$ and $\Pi_b$ -- are the lower and upper limits of the fundamental mode period.
Table~\ref{tabl1} also contains the characteristics of the period--luminosity relation
for the evolutionary sequence $Z=0.014$, $M_\mathrm{ZAMS}=1.5M_\odot$ (Fadeyev 2023).
As can be seen in Table~\ref{tabl1}, characteristics of this relation are intermediate
between those corresponding to stars with metallicities $Z=0.006$ and $Z=0.02$
for $M_\mathrm{ZAMS}=1.5M_\odot$.

\section*{conclusions}

Results of the present study based on consistent calculations of stellar evolution and
nonlinear stellar pulsations allow us to conclude that the metal abundance in Mira variables
substantially affects the period--luminosity relation.
The difference is most conspicuous in the period range $200~\textrm{day} < \Pi < 300~\textrm{day}$
where the luminosity of LMC Mira variables is higher than that of Galactic Mira variables
by 0.2 to 0.5 mag.
This is primarily due to an increase in the slope of the theoretical relation $k_L$ as
the metal abundance decreases by a factor of three from $Z=0.02$ to $Z=0.006$.
In particular, depending on the initial stellar mass $M_\mathrm{ZAMS}$ the change of the
coefficient $k_L$ ranges from $\approx 30\%$ to $\approx 50\%$
and this amount significantly exceeds the standard error of the slope coefficient
in the period--luminosity relation $\approx 6\%$ determined for $\approx 50$ Mira variables
of the LMC with periods $100~\textrm{day}\lesssim\Pi\lesssim400~\textrm{day}$
(Whitelock et al., 2008).
Therefore, one of the conclusions of the theory of stellar evolution and the theory of stellar
pulsation is that the slope of the period--luminosity relation of Galactic Mira variables is
significantly smaller in comparison that of Mira variables of the LMC.

To compare the theoretical period--luminosity relations with observations in more detail
one has to take into account the fact that the observed period--luminosity relation
is represented by stars with different initial masses.
To first approximation this can be taken into account by the use of weight coefficients
proportional to the initial mass function (Salpeter 1955).
Solution of this problem, however, requires much more evolutionary and pulsation computations.

\section*{references}

\begin{enumerate}

\item M.~Andriantsaralaza, S.~Ramstedt, W.H.T.~Vlemmings, and E.~De~Beck,
      Astron. Astrophys. \textbf{667}, A74 (2022).

\item Yu.A.~Fadeyev, Astron. Lett. \textbf{49}, 722 (2023).

\item Yu.A.~Fadeyev, Astron. Lett. \textbf{50}, 561 (2024).

\item M.W.~Feast, MNRAS \textbf{211}, 51 (1984).

\item M.W.~Feast, I.S.~Glass, P.A.~Whitelock, and R.M.~Catchpole, MNRAS \textbf{241}, 375 (1989).

\item M.W.~Feast, MNRAS \textbf{278}, 11 (1996).

\item I.S.~Glass and T.~Lloyd~Evans, Nature \textbf{291}, 303 (1981).

\item I.S.~Glass and M.W.~Feast, MNRAS \textbf{199}, 245 (1982).

\item S.M.G.~Hughes, Astron. J. \textbf{97}, 1634 (1989).

\item S.M.G.~Hughes and P.R.~Wood, Astron. J. \textbf{99}, 784 (1990).

\item P.~Iwanek, I.~Soszy\'nski, and S.~Koz\l{}owski, Astrophys. J. \textbf{919}, 99 (2021).

\item B.~Paxton, R.~Smolec, J.~Schwab, A.~Gautschy, L.~Bildsten, M.~Cantiello, A.~Dotter, R.~Farmer,
      J.A.~Goldberg, A.S.~Jermyn, S.M.~Kanbur, P.~Marchant, A.~Thoul, R.H.D.~Townsend, W.M.~Wolf,
      M.~Zhang, and F.X.~Timmes,
      Astrophys. J. Suppl. Ser. \textbf{243}, 10 (2019).

\item I.N.~Reid, S.M.G.~Hughes, and I.S.~Glass, MNRAS \textbf{275}, 331 (1995).

\item E.E.~Salpeter, Astrophys. J. \textbf{121}, 161 (1955).

\item J.L.~Sanders, MNRAS \textbf{523}, 2369 (2023).

\item I.~Soszy\'nski, W.A.~Dziembowski, A.~Udalski, M.~Kubiak, M.K.~Szyma\'nski,
      G.~Pietrzy\'nski, L.~Wyrzykowski, O.~Szewczyk O., and K.~Ulaczyk,
      Acta Astron \textbf{57}, 201 (2007).

\item E.~Vassiliadis and P.R.~Wood, Astrophys. J. \textbf{413}, 641 (1993).

\item P.A.~ Whitelock, M.W.~Feast, and F.~van Leeuwen, MNRAS \textbf{386}, 313 (2008).

\item P.R.~Wood and D.M.~Zarro, Astrophys. J. \textbf{247}, 247 (1981).
\item W.~Yuan, L.M.~Macri, S.~He, J.Z.~Huang, S.M.~Kanbur, and C.-C.~Ngeow, Astron. J. \textbf{154}, 149 (2017).

\end{enumerate}

\begin{table}[b]
\caption{Characteristics of period--luminosity relations.}
\label{tabl1}

\begin{center}
\begin{tabular}{cccccrr}
\hline
$Z$ & $M_\mathrm{ZAMS}/M_\odot$ & $k_L$ & $s_k$ & $a_0$ & $\Pi_a, \textrm{day}$ & $\Pi_b, \textrm{day}$ \\
\hline
0.006 & 1.1 &  0.966 &  0.03 &  1.28 & 124 & 256 \\
      & 1.3 &  0.925 &  0.11 &  1.43 & 185 & 287 \\
      & 1.5 &  1.163 &  0.10 &  0.89 & 187 & 316 \\
0.014 & 1.5 &  0.768 &  0.01 &  1.77 & 105 & 398 \\
0.020 & 1.1 &  0.737 &  0.04 &  1.62 & 127 & 394 \\
      & 1.3 &  0.706 &  0.02 &  1.80 &  78 & 416 \\
      & 1.5 &  0.713 &  0.01 &  1.86 &  68 & 389 \\
\hline
\end{tabular}
\end{center}

\end{table}

\newpage

\newpage

\end{document}